\documentstyle[twocolumn,pra,aps,epsfig]{revtex}
\begin{document}
\flushbottom
\draft
\title{Effects of atomic diffraction on the Collective Atomic Recoil Laser}
\author{M. G. Moore and P. Meystre}
\address{Optical Sciences Center and Department of Physics\\
University of Arizona, Tucson, Arizona 85721\\
(March 10, 1998)
\\ \medskip}\author{\small\parbox{14.2cm}{\small \hspace*{3mm}
We formulate a wave atom optics theory of the Collective Atomic Recoil Laser,
where the atomic center-of-mass motion is treated quantum mechanically.
By comparing the predictions of this theory with those of the ray atom
optics theory, which treats the center-of-mass atomic motion classically,
we show that for the case of a far off-resonant pump laser the
ray optics model fails to predict the linear response of the CARL
when the temperature is of the order of the recoil temperature
or less.  This is due to the fact that in this temperature regime 
one can no longer ignore the
effects of matter-wave diffraction on the atomic center-of-mass motion.
\\[3pt]PACS numbers: 42.55-f,42.50.Vk,03.75.-b  }}
\maketitle
\narrowtext

\section{Introduction}
The Collective Atomic Recoil Laser, or CARL, is the atomic equivalent of the
Free Electron Laser \cite{Bra90}. Developed theoretically by Bonifacio et al
\cite{BonSal94,BonSalNar94,BonSal95,SalCanBon95}, the CARL device
has three main components: (1) the active medium, which
consists of a gas of two-level atoms, (2) a strong pump
laser which drives the two-level atomic transition, and (3)
a ring cavity which supports an electromagnetic mode (the probe)
counterpropagating with respect to the pump.
Under suitable conditions, the operation of the CARL results in the
generation of a coherent light field (the probe) due to the following
mechanism. First, a weak probe field is initiated by noise, either optical
in the form of spontaneously emitted light, or atomic in
the form of density fluctuations in the atomic gas which backscatters the
pump. Once initiated, the probe combines with the pump field to form a
weak standing wave which acts as a periodic optical potential (light shift).
The center-of-mass motion of the atoms on this potential results in a
bunching (modulation) of their density, very much like the combined
effects of the wiggler and the light field leads to electron bunching in the
free-electron laser. This bunching process is then seen by the pump
laser as the appearance of a polarization grating in the active medium,
which results in stimulated backscattering into the probe field.
The resulting increase in the probe strength further increases the
magnitude of the standing wave field, resulting in more bunching
followed by an increase in stimulated backscattering, etc.
This positive feedback mechanism results in an exponential
growth of both the probe intensity and the atomic bunching.
This leads to the perhaps surprising result that the presence of the ring
cavity turns the ordinarily stable system of an atomic gas
driven by a strong pump laser into an unstable system.

The operation of the CARL was verified experimentally by Bigelow et al
\cite{HemBigKat96}, using a hot atomic cell.
Related experiments by Courtois et al \cite{CouGryLou93}  using
cold cesium atoms, and by Lippi et al \cite{LipBarBar96} using
hot sodium atoms measured the recoil induced small-signal probe gain,
which was interpreted in terms of coherent scattering off 
an induced polarization grating. However, these experiments lacked
a probe feedback mechanism, which is necessary to see the long time scale
instability which characterizes the CARL.

The CARL theory developed by Bonifacio et al
considers the atoms either as classical point particles moving in the
optical potential generated by the light fields, or, in a ``hybrid'' version,
as particles whose center-of-mass is labeled by their classical
position, but with quantum fluctuations about that position included.
From an atom optics point of view, such theories can be described as 
``ray atom optics'' treatments of the atomic field, in analogy with the 
ordinary ray optics treatment of electromagnetic fields.

Like ordinary ray optics, the ray atom optics description of CARL is
expected to be valid provided that the characteristic wavelength of the 
matter-wave field
remains much smaller than the characteristic length scale of any 
atom-optical element in the system. The characteristic wavelength of the
atomic field is its de Broglie wavelength, determined by the atomic mass and
the temperature $T$ of the atomic gas. The central atom-optical element of the
CARL is the periodic optical potential, which acts as a diffraction grating
for the atoms, and has the characteristic length scale of half the
optical wavelength. Hence the classical ``ray atom optics'' description
is intuitively expected to be valid provided that the temperature is high
enough that the thermal de Broglie wavelength is much smaller than the
optical wavelength. This gives the condition $T \gg T_R$, 
the recoil temperature of the
atoms, as the domain of ray atom optics. In particular, it is certainly expected to hold under the temperature
conditions of the experiments performed so far.

However, the spectacular recent progress witnessed by atomic cooling
techniques makes it likely that CARL experiments using ultracold atomic
samples can and will be performed in the future. In particular, subrecoil
temperatures can now be achieved almost routinely. The purpose of this
paper is to extend the CARL theory to this ``wave atom optics'' regime
\cite{MooMey98}.
In this regime matter-wave diffraction is expected to play a dominant role
in the CARL dynamics, and thus it becomes important to determine to what
extent it counteracts the bunching process in the CARL.

The wave optics theory of the CARL is similar to the analysis of atomic
diffraction by standing waves \cite{BerSho81}, except
that the electromagnetic field is now treated as a dynamical variable.
It is also similiar to the theory of recoil induced resonances
\cite{GuoBerDub92}, which describes the stimulated scattering of light off
a standing wave induced polarization grating, but the absence of a feedback
mechanism for the probe feedback in that case means that it lacks the
instability necessary for lasing.

In this paper we focus on the case of a far off-resonant pump laser,
thus permitting us to neglect the excited state population and
therefore to ignore the effects of spontaneous emission (except as
a hypothetical source of noise for probe initialization).
We further concentrate on the linear regime, where both the probe field
and the atomic bunching are considered as infinitesimal quantities,
since it is this regime that determines whether or not the exponential
instability occurs. Finally, we restrict our analysis to
atomic densities low enough that collisions between atoms
may be ignored, and neglect the transverse motion of the atoms, which in the
absence of collisions is decoupled from the longitudinal degree of freedom
along which bunching occurs.

We note at the outset that our theory is semiclassical in that it treats
the electromagnetic field classically. While this approximation can not 
fully describe the statistical properties of the CARL output, 
it is sufficient to describe the small-signal gain of the system, 
provided that one makes the implicit assumption that small fluctuations will
trigger it, an approach familiar from conventional laser theory and
nonlinear optics. We 
also emphasize that it is not inconsistent to treat the matter waves quantum
mechanically while treating the light classically, since the limits under which 
a quantum
description is required are independent. For light, this limit is usually
associated with weak intensities, while for matter waves it is normally a low
temperature limit.

This rest of this paper is organized as follows: Section II briefly reviews
the ray atom optics model of the CARL, establishing the notation and setting
the stage for a comparison of its predictions with those of the wave atom
optics theory, which is introduced in section III. Section IV discusses the
collective instability leading to CARL operation, compares the ray atom
optics and the wave atom optics predictions, and determines the domain of
validity of the former theory. Finally section IV is a summary and outlook.

\section{Ray Atom Optics model}
The Ray Atom Optics (RAO) model of the CARL has been developed and extensively
studied by Bonifacio et al. \cite{BonSal94,BonSalNar94,BonSal95,SalCanBon95}
It begins with the classical $N$-particle Hamiltonian
\begin{equation}
H_N=\sum_{j=1}^NH_1(z_j,p_j),
\label{HN}
\end{equation}
where $z_j$ and $p_j$ are the classical position and momentum of the
$j$th atom, obeying the canonical equations of motion
$dz_j/dt=\partial H_N/ \partial p_j$ and $dp_j/dt=-\partial H_N/\partial z_j$.
The single-particle Hamiltonian $H_1$ is given explicitly by
\begin{eqnarray}
H_1(z_j,p_j)&=&\frac{p_j^2}{2m}+\frac{\hbar\omega_0}{2}\sigma_{zj}
+i\hbar\left[g_1a_1^\ast
e^{-ik_1z_j}\sigma_{-j}
\right.\nonumber\\
&+&\left.g_2a_2^\ast e^{-ik_2z_j}\sigma_{-j}
-c.c.\right],
\label{H1}
\end{eqnarray}
where $m$ is the atomic mass, $\omega_0$ is the natural frequency of the atomic
transition being driven by the pump and probe lasers, and $g_1$ is the
atom-probe electric dipole coupling constant. It is
given by $g_1=\mu_1[ck_1/(2\hbar\epsilon_0V)]^{1/2}$, where $\mu_1$ is
the projection of the atomic dipole moment along the probe polarization,
$k_1$ is the probe wavenumber, and $V$ is  the quantization volume.
The atom-pump coupling constant
$g_2$ is defined analogously to $g_1$,
but depending on $\mu_2$, the projection of the atomic dipole moment along
the pump polarization, and $k_2$ the pump wavenumber.
The normal variables $a_1$ and $a_2$ describe the probe and pump laser fields,
respectively.  They obey Maxwell's equation
\begin{equation}
\frac{d}{dt}a_i=-i\omega_ia_i+g_i\sum_{j-1}^Ne^{-ik_iz_j}\sigma_{-j},
\label{maxwell}
\end{equation}
where $\omega_i$ is the natural frequency of the probe $(i=1)$ or the pump 
$(i=2)$ 
field. Note that these equations are also valid for quantized
electromagnetic fields, provided that $a_i$ are interpreted as annihilation
operators, but we describe the light fields classically in this paper.

The variables $\sigma_{-j}$ and $\sigma_{zj}$ are the expectation values
of the quantum mechanical Pauli pseudo-spin operators which describe the
internal state of the $j$th atom.  They obey the familiar optical Bloch
equations, appropriately modified to include the center-of-mass motion of
the atoms and with spontaneous emission neglected
\footnote{Spontaneous emission is neglected in anticipation of the future
approximation that the pump lasers are far-off resonant, and therefore the
excited state population may be safely neglected.},
\begin{equation}
\frac{d}{dt}\sigma_{-j}=-i\omega_0\sigma_{-j}+\left[g_1^\ast a_1e^{ik_1z_j}
+g_2^\ast a_2e^{ik_2z_j}\right]\sigma_{zj},
\label{bloch2}
\end{equation}
and
\begin{equation}
\frac{d}{dt}\sigma_{zj}=-2\left[g_1a_1^\ast e^{-ik_1z_j}
+g_2a_s^\ast e^{-ik_2z_j}\right]\sigma_{-j}+c.c.
\label{bloch1}
\end{equation}

It is convenient to introduce slowly varying variables via the transformations
$a_1=a_1^\prime e^{-i\omega t}$, $a_2=a_2^\prime e^{-i\omega t}$, and
$\sigma_{-j}=\sigma_{-j}^\prime e^{-i(\omega t-k_2z_j)}$, where $\omega$
is the pump frequency shifted by the frequency pulling contribution due
to the induced atomic polarization. The exact value of $\omega$ will be derived
in a self-consistent manner shortly in a way similar to the approach used in
conventional laser theory. These new variables obey the
equations of motion
\begin{equation}
\frac{d}{dt}z_j=\frac{p_j}{m},
\label{rotz}
\end{equation}
\begin{eqnarray}
\frac{d}{dt}p_j&=&-\hbar\left[g_1 k_1 {a_1^\prime}^\ast e^{-i(k_1-k_2)z_j}
+g_2 k_2{a_2^\prime}^\ast\right]\sigma_{-j}^\prime\nonumber\\
&+&c.c.,
\label{rotp}
\end{eqnarray}
\begin{equation}
\frac{d}{dt}a_1^\prime=i(\omega-\omega_1)a_1^\prime+g_1\sum_{j=1}^N
e^{-i(k_1-k_2)z_j}\sigma_{-j}^\prime,
\label{rota1}
\end{equation}
\begin{equation}
\frac{d}{dt}a_2^\prime=i(\omega-\omega_2)a_2^\prime+g_2\sum_{j=1}^N
\sigma_{-j}^\prime,
\label{rota2}
\end{equation}
\begin{eqnarray}
\frac{d}{dt}\sigma_{zj}&=&-2\left[g_1{a_1^\prime}^\ast e^{-i(k_1-k_2)z_j}
+g_2{a_2^\prime}^\ast\right]\sigma_{-j}^\prime\nonumber\\
&+&c.c.,
\label{rotsigmaz}
\end{eqnarray}
and
\begin{eqnarray}
\frac{d}{dt}\sigma_{-j}^\prime&=&i(\omega-\omega_0-\frac{k_2}{m}p_j)
\sigma_{-j}^\prime\nonumber\\
&+&\left[g_1^\ast a_1^\prime e^{i(k_1-k_2)z_j}
+g_2^\ast a_2^\prime\right]\sigma_{zj}.
\label{rotsigma-}
\end{eqnarray}

In the case where the lasers are tuned far off resonance, and the atoms
are initially in the ground state, the excited state population remains
small and can be neglected.  This is equivalent to describing the
atoms as classical Lorentz atoms, and is accomplished by setting
$\sigma_{zj}=-1$ in Eq. (\ref{rotsigma-}). Assuming further that the
detuning $\omega-\omega_0$ is much larger than any other frequency
in Eq. (\ref{rotsigma-}), allows one to adiabatically eliminate
$\sigma_{-j}^\prime$ with
\begin{equation}
\sigma_{-j}^\prime\approx-\frac{i}{(\omega-\omega_0)}\left[g_1^\ast
a_1^\prime e^{i(k_1-k_2)z_j}+g_2^\ast a_2^\prime\right],
\label{adelim}
\end{equation}
where we have in addition neglected the Doppler shift $k_2p_j/m$ compared
to $\omega-\omega_0$. This leads to the reduced set of equations
\begin{equation}
\frac{d}{dt}z_j=\frac{p_j}{m},
\label{adelz}
\end{equation}
\begin{equation}
\frac{d}{dt}p_j=-i\frac{2\hbar k_0}{(\omega-\omega_0)}\left[
g_1^\ast g_2 {a_2^\prime}^\ast a_1^\prime e^{i2k_0z_j}-c.c.\right],
\label{adelp}
\end{equation}
\begin{eqnarray}
\frac{d}{dt}a_1^\prime&=&i\left[\omega-\frac{N|g_1|^2}{(\omega-\omega_0)}
-\omega_1\right]a_1^\prime\nonumber\\
&-&i\frac{g_2^\ast g_1}{(\omega-\omega_0)}a_2^\prime
\sum_{j=1}^Ne^{-i2k_0z_j},
\label{adela1}
\end{eqnarray}
and
\begin{eqnarray}
\frac{d}{dt}a_2^\prime&=&i\left[\omega-\frac{N|g_2|^2}{(\omega-\omega_0)}
-\omega_2\right]a_2^\prime\nonumber\\
&-&i\frac{g_1^\ast g_2}{(\omega-\omega_0)}a_1^\prime
\sum_{j=1}^Ne^{i2k_0z_j},
\label{adela2}
\end{eqnarray}
where we have introduced $k_0=(k_1-k_2)/2$.

We now introduce the undepleted pump approximation, valid in the linear
regime where $a_1^\prime$ remains small.  This is achieved by dropping the
term proportional to $a_1^\prime$  in Eq. (\ref{adela2}).  This yields
\begin{equation}
\frac{d}{dt}a_2^\prime=i\left[\omega-\frac{N|g_2|^2}{(\omega-\omega_0)}-
\omega_2\right]a_2^\prime,
\label{undepa2}
\end{equation}
which has the steady state solution $a_2^\prime(t)=a_2(0)$ provided that the
frequency pulling condition
\begin{equation}
\omega-\frac{N|g_2|^2}{(\omega-\omega_0)}-\omega_2=0.
\label{steadyst}
\end{equation}
is satisfied. Note that this equation has two solutions, but we must
choose the branch which gives the result $\omega=\omega_2$ when $N=0$.
This leads to the solution
\begin{equation}
\omega=\frac{1}{2}\left[\omega_0+\omega_2\pm\sqrt{(\omega_2-\omega_0)^2
+4N|g_2|^2}\right],
\label{omega}
\end{equation}
where the plus sign must be taken for positive detunings
$(\omega_2>\omega_0)$ and the minus sign for negative detunings
$(\omega_2<\omega_0)$. Expanding this relation to lowest order
in $(\omega_2-\omega_0)^{-1}$ gives the expected result
\begin{equation}
\omega\approx\omega_2+\frac{N|g_2|^2}{(\omega_2-\omega_0)}.
\label{expandomega}
\end{equation}

To proceed analytically past this point, it is convenient to introduce
the dimensionless variables $\theta_j \equiv 2k_0z_j$, $P_j=p_j/\hbar k_0$,
$A=g_1^\ast g_2a_2^\ast(0)a_1^\prime/[\omega_r(\omega-\omega_0)]$
and $\tau=4\omega_r t$,
where the recoil frequency $\omega_r$ is given by
\begin{equation}
\omega_r=\hbar k_0^2/2m.
\end{equation}
These variables obey the equations of motion
\begin{equation}
\frac{d}{d\tau}\theta_j=P_j,
\label{dimtheta}
\end{equation}
\begin{equation}
\frac{d}{d\tau}P_j=-i A e^{i\theta_j}+c.c.,
\label{dimP}
\end{equation}
and
\begin{equation}
\frac{d}{d\tau} A=i\Delta A
-i\alpha\frac{1}{N}\sum_{j=1}^Ne^{-i\theta_j},
\label{dimdA}
\end{equation}
where we have introduced the dimensionless control parameters
\begin{equation}
\Delta=\delta_1/4\omega_r,
\label{defDelta}
\end{equation}
and
\begin{equation}
\alpha=N|g_1|^2|g_2|^2|a_2(0)|^2/8\omega_r^2(\omega-\omega_0)^2,
\label{defalpha}
\end{equation}
where $\delta_1=\omega-\omega_1-N|g_1|^2/(\omega-\omega_0)$.
We note that both $\Delta$ and $\alpha$ are real numbers, and
furthermore that $\alpha\ge 0$.

We seek solutions of these equations which are perturbations about the case
$A=0$. Thus we make the substitutions
\begin{equation}
\theta_j=\theta_j(0)+P_j(0)\tau+\delta\theta_j,
\label{perttheta}
\end{equation}
and
\begin{equation}
P_j=P_j(0)+\delta P_j,
\label{pertP}
\end{equation}
where $\theta_j(0)$ is randomly taken from a uniform distribution, and
$P_j(0)$ is randomly taken from the initial momentum distribution.
The new variables $\delta\theta_j$ and $\delta P_j$ give the perturbations
on the atomic center-of-mass motion
due to a nonzero $A(0)$. 
We introduce finally the linearized velocity group bunching parameter and its
``conjugate'' momentum according to
\begin{equation}
B(k)=\frac{1}{N}\sum_{j=1}^N\delta_{P_j(0),P(k)}(1-i\delta\theta_j)
e^{-i(\theta_j(0)+P_j(0)\tau)},
\label{vgbunch}
\end{equation}
and
\begin{eqnarray}
{\mit\Pi}(k)&=&\frac{1}{N}\sum_{j=1}^N\delta_{P_j(0),P(k)}\delta P_j
e^{-i(\theta_j(0)+P_j(0)\tau)}\nonumber\\
&+&P(k)B(k).
\label{vgmom}
\end{eqnarray}
We note that
\begin{equation}
\sum_k B(k)=\langle e^{-i2k_0z}\rangle,
\label{bunch}
\end{equation}
and the amplitude of (\ref{bunch}) is a measure of the degree of bunching of
the atomic gas.  A magnitude of zero indicates no bunching, while
a magnitude of one indicates maximum bunching. This leads to the equations
\begin{equation}
\frac{d}{d\tau}B(k)=-i{\mit\Pi}(k),
\label{dBdt}
\end{equation}
\begin{equation}
\frac{d}{d\tau}{\mit\Pi}(k)=i\left[P^2(k)B(k)-2P(k){\mit\Pi}(k)-
\frac{N(k)}{N}A\right],
\label{dPidt}
\end{equation}
and
\begin{equation}
\frac{d}{d\tau}A=i\left[\Delta A-\alpha\sum_k B(k)\right],
\label{dAdt}
\end{equation}
where $N(k)$ is the number of atoms in the velocity group with momentum
$\hbar k_0 P(k)$ and we have assumed that
\begin{equation}
\sum_{j=1}^N\delta_{P_j(0),P(k)}e^{-i2\theta_j(0)}=0,
\label{novgb}
\end{equation}
an assumption that requires that $N(k)\gg 1$. Note that this formulation
implies a discretization of the initial momentum distribution, and furthermore
assumes that the atomic positions in each velocity group are initially randomly
distributed along the CARL cavity. Fluctuations in the initial distributions
can of course readily be included into the initial conditions of the
perturbation variables.

\section{Wave atom optics model}
In order to quantize the center-of-mass motion of a gas of Bosonic
atoms, one may either utilize first quantization, and replace the variables
$z_j$ and $p_j$ in the $N$-particle Hamiltonian (\ref{HN}) with operators
satisfying the canonical commutation relations $[\hat{z}_j,
\hat{p}_{j^\prime}]=i\hbar\delta_ {jj^\prime}$, or equivalently we 
can second-quantize the single particle Hamiltonian (\ref{H1}), introducing 
creation and annihilation operators for excited and ground state atoms of a
given center-of-mass momentum. It is this second method which
we will adopt in deriving the Wave Atom Optics (WAO) model. 
In the absence of collisions, the
second-quantized Hamiltonian is simply
\begin{equation}
\hat{H}=\sum_k \hat{H}(k),
\label{hatH}
\end{equation}
where $\hat{H}(k)$ is given by
\begin{eqnarray}
\hat{H}(k)&=&\frac{\hbar^2 k^2}{2m}\hat{c}^\dag_g(k)\hat{c}_g(k)
+\left(\frac{\hbar^2k^2}{2m}+\hbar\omega_0\right)
\hat{c}^\dag_e(k)\hat{c}_e(k) \nonumber\\
&+&i\hbar\left[g_1a_1^\ast\hat{c}^\dag_g(k+k_1)\hat{c}_e(k)
+g_2a_2^\ast\hat{c}^\dag_g(k+k_2)\hat{c}_e(k)\right.\nonumber\\
&-&\left. H.c.\right],
\label{Hk}
\end{eqnarray}
where the field operator $\hat{c}_g(k)$ annihilates
a ground state atom of momentum $\hbar k$, and $\hat{c}_e(k)$
annihilates an excited atom of momentum $\hbar k$.  We assume that the 
atoms in the sample are bosonic, so that these operators
obey the commutation relations
\begin{equation}
[\hat{c}_g(k),\hat{c}^\dag_g(k^\prime)]=
[\hat{c}_e(k),\hat{c}^\dag_e(k^\prime)]=\delta_{kk^\prime},
\label{comm}
\end{equation}
all other commutators being equal to zero.

With the atomic polarization now expressed in terms of
field operators, Maxwell's equations (\ref{maxwell})
for the classical laser fields become
\begin{equation}
\frac{d}{dt}a_i=-i\omega_ia_i+g_i\sum_k\langle
\hat{c}^\dag_g(k+k_i)\hat{c}_e(k)\rangle.
\label{maxwell2}
\end{equation}
Hence, all that is required to determine the field evolution are the
expectation value of bilinear combinations of atomic creation and
annihilation operators. The evolution of these expectation values is easily
obtained by introducing the ``single-particle'' atomic density operators
\footnote {These are single-particle operators in the sense of many-body
theory, since they only involve the annihilation of an atom in a given
state and its creation in some other state.}
\begin{equation}
\hat{\rho}_{gg}(k,k^\prime)=\hat{c}^\dag_g(k^\prime)\hat{c}_g(k),
\label{rhogg}
\end{equation}
\begin{equation}
\hat{\rho}_{eg}(k,k^\prime)=[\hat{\rho}_{ge}(k^\prime,k)]^\dag=
\hat{c}^\dag_g(k^\prime)\hat{c}_e(k),
\label{rhoeg}
\end{equation}
and
\begin{equation}
\hat{\rho}_{ee}(k,k^\prime)=\hat{c}^\dag_e(k^\prime)\hat{c}_e(k).
\label{rhoee}
\end{equation}
Note that e.g. the expectation value of the diagonal operator 
$\langle \hat{\rho}_{gg}(k,k)\rangle$ gives the
mean number of ground state atoms with momentum $\hbar k$.
The expectation values of these operators obey the equations of motion
\begin{equation}
\frac{d}{dt}\rho_{jj^\prime}(k,k^\prime)=\frac{i}{\hbar}\langle[\hat{H},
\hat{\rho}_{jj^\prime}(k,k^\prime)]\rangle
\label{Heisrho}
\end{equation}
where
$\rho_{jj^\prime}(k,k^\prime)=\langle\hat{\rho}_{jj^\prime}(k,k^\prime)\rangle$.
The full form of these equations is given in the Appendix. The important
point is that they are depend only on $\rho_{jj^\prime}(k,k^\prime)$,
hence they form a closed set of equations which describe the response of the
atomic field to the driving laser fields. 
We note that had we included collisions in our model, this would no longer be
the case.

Introducing in analogy to the ray optics description the rotating variables
$a_1=a_1^\prime e^{-i\omega t}$, $a_2=a_2^\prime e^{-i\omega t}$, and
$\rho_{eg}(k,k^\prime)=\rho_{eg}^\prime(k-k_2,k^\prime)e^{-i\omega t}$,
neglecting the excited state population, and solving adiabatically for
$\rho_{eg}^\prime(k,k^\prime)$ yields
\begin{eqnarray}
\rho_{eg}^\prime(k,k^\prime)&\approx&-
\frac{i}{(\omega-\omega_0)}[g^\ast_1a_1^\prime
\rho_{gg}(k+2k_0,k^\prime)\nonumber\\
&+&g^\ast_2a_2^\prime\rho_{gg}(k,k^\prime)].
\label{adelimeg}
\end{eqnarray}
Substituting Eq. (\ref{adelimeg}) into Maxwell's equation 
(\ref{maxwell2}) for the pump
and making once more the undepleted pump approximation leads to the solution
$a^\prime_2(t)=a_2(0)$
provided that $\omega$ is given by Eq. (\ref{omega}).
We then substitute Eq. (\ref{adelimeg}) into the equation of motion for
$\rho_{gg}(k,k^\prime)$, and introduce the dimensionless wavenumber 
$\kappa=k/(k_1-k_2)$ and the mean density $\rho(\kappa,\kappa^\prime)=
\rho_{gg}(k,k^\prime)/N$, in addition to the dimensionless variables
already defined in the ray atom optics model. We arrive at the 
wave optics equations of motion
\begin{eqnarray}
\frac{d}{d\tau}\rho(\kappa,\kappa^\prime)&=&-i(\kappa^2-{\kappa^\prime}^2)
\rho(\kappa,\kappa^\prime)\nonumber\\
&+&\frac{i}{2}A^\ast
[\rho(\kappa,\kappa^\prime+1)-\rho(\kappa-1,\kappa^\prime)]\nonumber\\
&-&\frac{i}{2}A
[\rho(\kappa+1,\kappa^\prime)-\rho(\kappa,\kappa^\prime-1)],
\label{drhodtau}
\end{eqnarray}
and
\begin{equation}
\frac{d}{d\tau}A=i\Delta A-i\alpha\sum_\kappa
\rho(\kappa,\kappa+1),
\label{dAdtau}
\end{equation}
where the parameters $\Delta$ and $\alpha$ are given by Eqs. (\ref{defDelta})
and (\ref{defalpha}), respectively.

As in Sec. II, we seek a solution which is a perturbation about the
case $A=0$. From Eq. (\ref{drhodtau}), the unperturbed solution is readily
found to be
\begin{equation}
\rho(\kappa,\kappa^\prime,\tau)=\rho(\kappa,\kappa^\prime,0)
e^{-i(\kappa^2-{\kappa^\prime}^2)\tau}.
\label{solution1}
\end{equation}
We consider specifically an atomic sample initially in thermal equilibrium,
so that Eq. (\ref{solution1}) becomes
\begin{equation}
\rho(\kappa,\kappa^\prime,\tau)=\frac{N(\kappa)}{N}
\delta_{\kappa,\kappa^\prime},
\label{unpertrho}
\end{equation}
where $N(\kappa)$, the number of atoms with initial wavenumber
$2k_0\kappa$, is given by a thermal distribution function. We introduce the
perturbation variables
$\delta\rho(\kappa,\kappa^\prime)$ according to
\begin{equation}
\rho(\kappa,\kappa^\prime)=\frac{N(\kappa)}{N}\delta_{\kappa,\kappa^\prime}
+\delta\rho(\kappa,\kappa^\prime).
\label{fluctrho}
\end{equation}
and observe that 
Maxwell's equation (\ref{dAdtau}), which becomes
\begin{equation}
\frac{d}{d\tau}A=i\Delta A -i\alpha\sum_k\delta\rho(\kappa,\kappa+1),
\label{A}
\end{equation}
together with the linearized equation
\begin{eqnarray}
\frac{d}{d\tau}\delta\rho(\kappa,\kappa+1)&=&
i(2\kappa+1)\delta\rho(\kappa,\kappa+1)\nonumber\\
&-&i\frac{[N(\kappa+1)-N(\kappa)]}{2N}A .
\label{drho}
\end{eqnarray}
form a closed set of equations which underlies the dynamics of the 
CARL in the linear regime of wave atom optics.

\section{Collective instability}

The most important feature of the CARL is the appearance of a collective
instability, which gives rise to exponential gain under appropriate parameter
settings.  This instability is characterized by an imaginary frequency component
in the spectrum of the probe field $A(\tau)$. As has been demonstrated in
Ref. \cite{BonSal95}, one needs not solve the complete set of equations derived
in the previous sections in order to determine the necessary conditions for
the collective instability.  Instead, by taking the Laplace transform of 
these equations one can derive a ``characteristic equation'' which allows 
one to determine whether exponential gain occurs, and if so what the 
exponential growth rate is.

For the Ray Atom Optics model, the Laplace transform of Eq.
(\ref{dAdt}) yields
\begin{equation}
\tilde{A}_R(s)=\frac{A(0)}{R(s)},
\label{raolapA}
\end{equation}
where $R(s)$ is given by
\begin{equation}
R(s)=\left[s-i\Delta-i\alpha\int\frac{f(k)dk}{(s+i2k)^2}\right].
\label{raoce}
\end{equation}
In obtaining this result we have taken the continuum limit and assumed 
that $B(k)$ and ${\mit\Pi}(k)$ vanish at $\tau=0$. Here $f(k)$ is simply 
the normalized thermal distribution function for the dimensionless 
momentum $P(k)=2k$. The roots of $R(s)$ give the characteristic exponents 
of the CARL. Stability requires that all roots be purely imaginary. 
When the collective instability occurs, however, there will be
one root with a positive 
real part. This real part is the RAO exponential growth rate $\Gamma_R$.
This result is identical to that first obtained by Bonifacio et al.

The Wave Atom Optics model, which includes the effects 
of atomic diffraction, yields the Laplace transform
\begin{equation}
\tilde{A}_W(s)=\frac{A(0)}{W(s)}.
\label{waolapA}
\end{equation}
$W(s)$ is given by
\begin{equation}
W(s)=\left[s-i\Delta
-i\alpha\int\frac{f(k)dk}
{(s-i(2k-1))(s-i(2k+1))}\right],
\label{waoce}
\end{equation}
where we have again taken the continuum limit and assumed that
$\delta\rho(k,k+1)$ vanishes at $\tau=0$.
If a root of $W(s)$ with a positive real part exists, that
real part is the WAO exponential growth rate $\Gamma_W$.

We see by comparing Eqs. (\ref{raolapA}) and (\ref{waolapA})
that the effect of atomic diffraction is to lift the degeneracy of the 
singularity under the integral. This expression also leads us immediately 
to the conclusion that if the width of the momentum distribution
$f(k)$ is large compared to $2k$, then the singularity will appear 
as essentially degenerate, and the effects of matter waves diffraction will
be negligible, Thus the RAO and WAO models should agree for
large enough temperatures.

\subsection{Finite temperatures}

In the absence of quantum degeneracies, the thermal momentum distribution 
is given by the Maxwell-Boltzmann distribution
\begin{equation}
f(k)=\frac{2\beta}{\sqrt{\pi}}e^{-4k^2\beta^2},
\label{thermal}
\end{equation}
where $\beta^2=T_R/T$ and $T_R=\hbar\omega_r/k_B$ is the recoil temperature,
$K_B$ being the Boltzmann constant.
By substituting Eq. (\ref{thermal}) into Eq.  (\ref{raoce}) and using the 
Fourier convolution theorem we find that the RAO exponential growth rate 
$\Gamma_R$ is determined by the equation
\begin{equation}
s-i\Delta-i\alpha\int_0^\infty pe^{-p^2/4\beta^2-ps}dp=0,
\label{raoconv}
\end{equation}
which can be integrated to give the transcendental equation
\begin{equation}
s-i\Delta-i2\alpha\beta^2+i2\sqrt{\pi}\alpha\beta^3e^{\beta^2s^2}
\mbox{erfc}(\beta s)=0.
\label{raoce2}
\end{equation}
In contrast, substituting Eq. (\ref{thermal}) into Eq. (\ref{waoce}) and again
using the convolution theorem we find that the WAO exponential growth rate
$\Gamma_W$ is determined by equation
\begin{equation}
s-i\Delta-i\alpha\int_0^\infty e^{-p^2/4\beta^2-ps}\sin(p)=0,
\label{waoconv}
\end{equation}

By examining Eq. (\ref{waoconv}) we see that in the case $\beta \ll 1$
we are justified in expanding $\sin(p)$ to
lowest order in $p$. This exactly reproduces Eq. (\ref{raoconv}), thus showing
that the WAO and RAO descriptions make indistinguishable
predictions about the exponential growth rate in the limit
$T\gg T_R$.
However, for temperatures comparable to or less than
the recoil temperature, we will see that the RAO theory fails to correctly 
predict
the behavior of the CARL in the linear regime.  Physically, this is due
to the fact that it does not account for the effects of atomic diffraction,
which tends to counteract the bunching process. Finally, we note that
upon integration, Eq. (\ref{waoconv}) becomes the
transcendental equation
$$
s-i\Delta+\frac{\sqrt{\pi}}{2}\alpha\beta e^{\beta^2(s^2-1)}
\left[e^{i2\beta^2s}\mbox{erfc}[\beta(s+i)]\right.
$$
\begin{equation}
\left.-e^{-i2\beta^2s}\mbox{erfc}[\beta(s-i)]\right]=0.
\label{waoce2}
\end{equation}
In the next subsection we will examine in more detail the precise manner
in which diffraction interferes 
with the bunching process for the special case of a zero
temperature atomic gas. But before turning to this extreme situation,
we present numerical results comparing RAO and WAO models
at non-zero temperature, as determined by solving Eqs. (\ref{raoce2})
and (\ref{waoce2}).

Figures 1(b-d) compare $\Gamma_R$ with $\Gamma_W$ at $\alpha=10$ for the
three different temperature regimes, $T=T_R$, $T=10T_R$, and $T=100T_R$
respectively. Figures 2(b-d) shows the same comparison for $\alpha=10^{-1}$.
While we see that the behavior of $\Gamma_R$ and $\Gamma_W$ depends strongly
on $\alpha$ (recall that $\alpha$ is 
proportional to both the pump intensity and
the atomic density), the discrepancies between the two models
as a function of temperature are very similar. At $T=T_R$
there are significant differences between the predictions of the RAO and WAO
models, but these differences become minimal at $T=10T_R$, and
insignificant at $T=100T_R$. We also observe that the differences are
more pronounced for lower values
of $\alpha$, meaning that at lower densities and/or pump intensities,
the quantum mechanical behavior becomes more apparent. The reason for
this is that at high intensities the bunching process, driven by the
probe field, dominates, while at low intensities the anti-bunching
effects of atomic diffraction play a larger role.

\subsection{The $T=0$ limit}
For a typical atom, the recoil temperature is of the order of microkelvins,
e.g. for sodium we have $T_R=2.4\mu K$. However, recent advances in 
cooling techniques have led to measured temperatures as low as the picokelvin
regime.  At these extreme temperatures the condition
$T\ll T_R$ is satisfied, i.e. we are effectively in the $T\to 0$ limit .
In this section we study the $T=0$ case in detail in order to gain further 
insight into the exact role of matter wave diffraction in the CARL system.

For the RAO model, we have a single velocity group at $k=0$. 
Thus by differentiating Eq. (\ref{dBdt}) with respect to $\tau$ and using
Eq. (\ref{dPidt}), we see that the bunching parameter $B\equiv B(0)=
\langle\exp(-i2k_0z)\rangle$ obeys the equation of motion
\begin{equation}
\frac{d^2}{d\tau^2}B=-A,
\label{zeroBrao}
\end{equation}
where we have taken $P(0)=0$ and $N(0)=N$ to indicate that all atoms are
initially at rest.

In the WAO description, setting $N(\kappa)/N=\delta_{\kappa,0}$
in Eq. (\ref{drho}) shows that two variables are coupled to
the probe field, $\delta\rho(-1,0)$, and $\delta\rho(0,1)$. They describe
the recoil of atoms initially at rest as a result of their interaction
with the light fields. We proceed then by introducing the new variable 
$B \equiv \delta\rho(-1,0)+\delta\rho(0,1)$, which
has the same physical meaning as in the RAO model, namely
$B=\langle\exp(-i2k_0z)\rangle$. But in contrast to that case, the time
evolution of $B$ is
now governed by the equation of motion
\begin{equation}
\frac{d^2}{d\tau^2}B=-B-A.
\label{zeroBwao}
\end{equation}
This result shows that in contrast to the predictions of classical mechanics,
where the bunching parameter $B$ has dynamics similar to a {\it free
particle} driven by the probe field $A$, quantum mechanically $B$ behaves 
as a {\it simple harmonic oscillator} of frequency $4\omega_r$ 
(in original time units), and subject to that same driving force.
In the linear regime, $B$ is assumed to be a small perturbation about
its initial value of zero, and the forces resulting from a non-zero probe
field $A$ tend to cause $B$ to increase.  But this mechanism is opposed by
the ``restoring force'' due to atomic diffraction.
\centerline{\psfig{figure=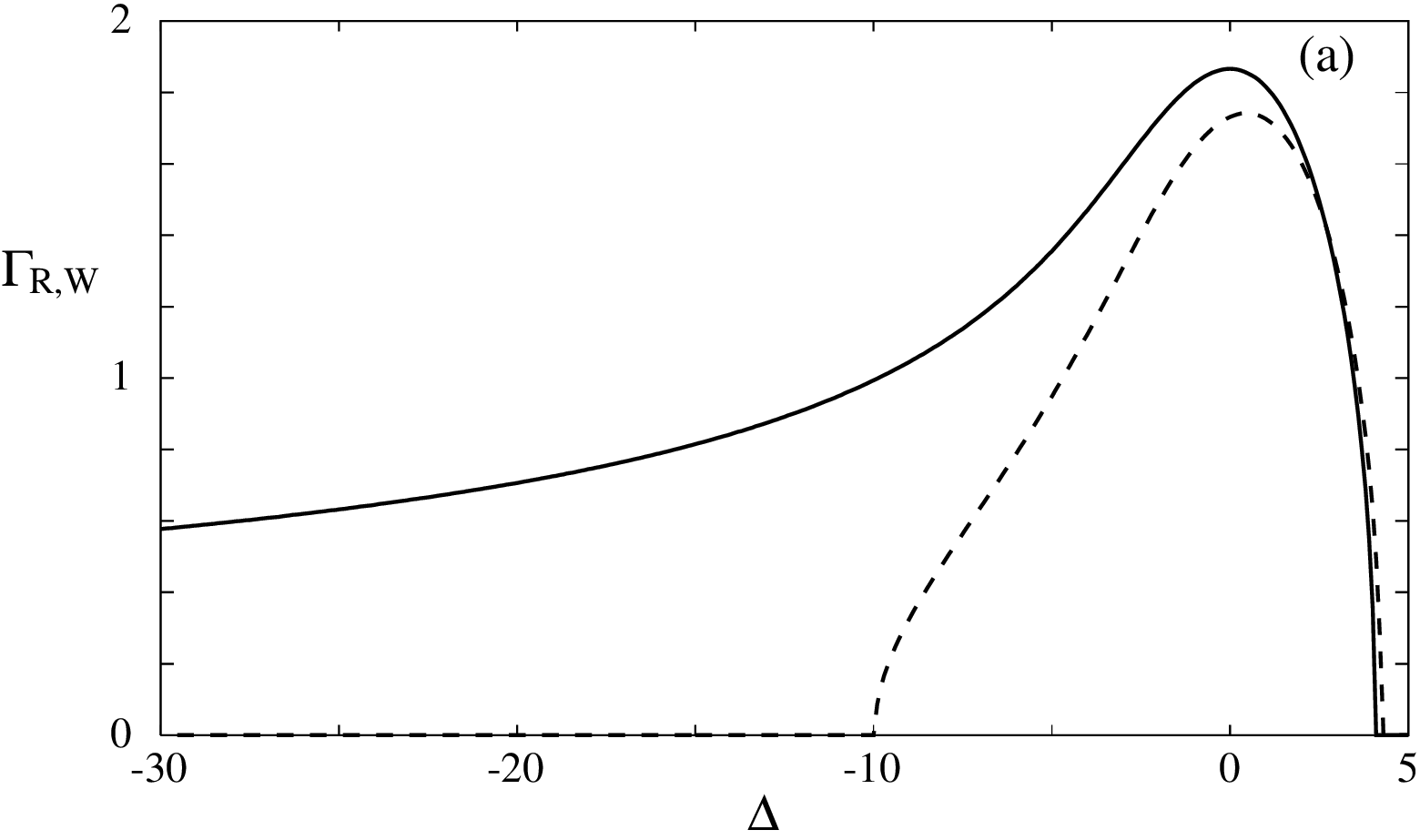,width=8.6cm,clip=}}
\centerline{\psfig{figure=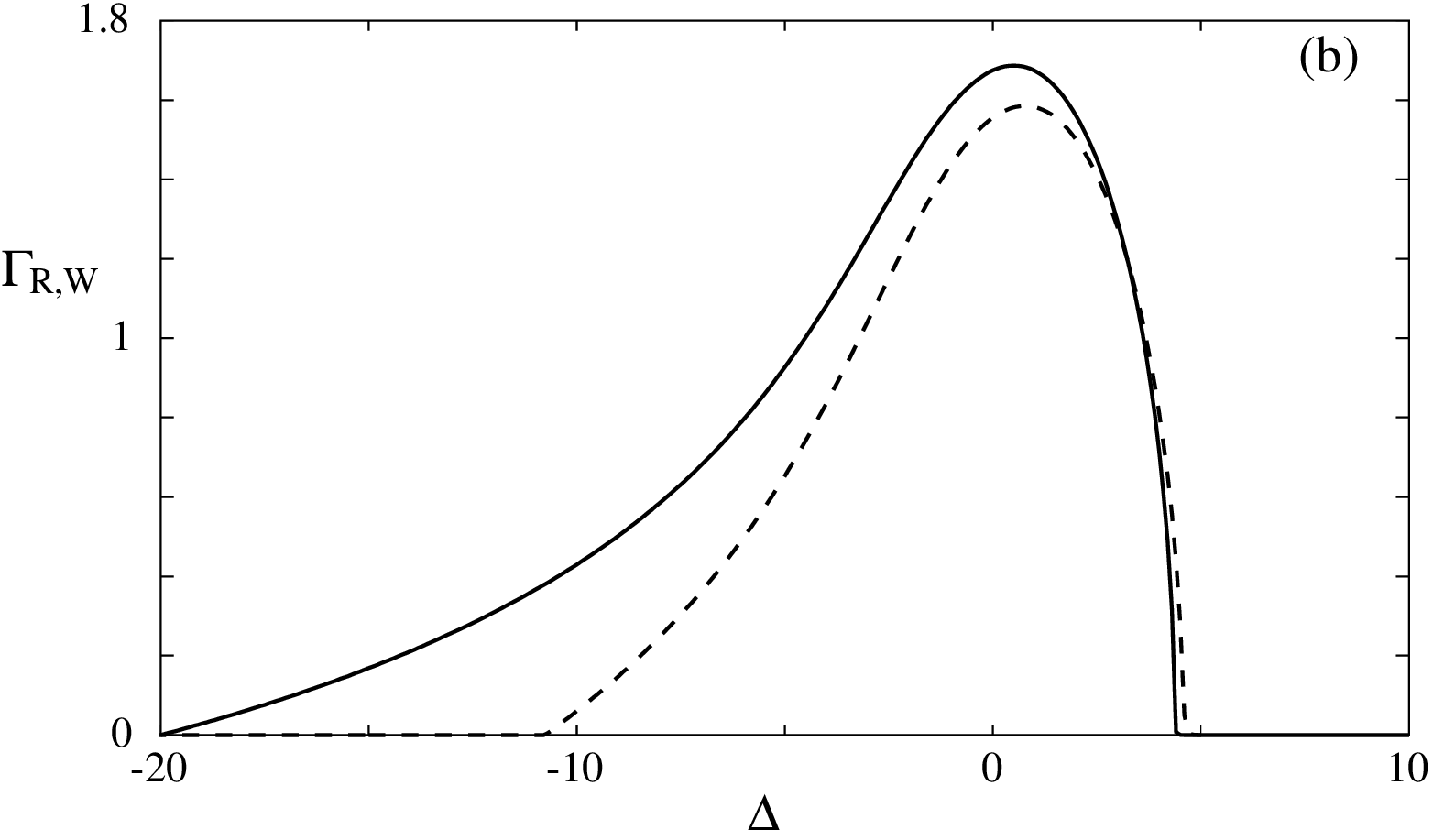,width=8.6cm,clip=}}
\centerline{\psfig{figure=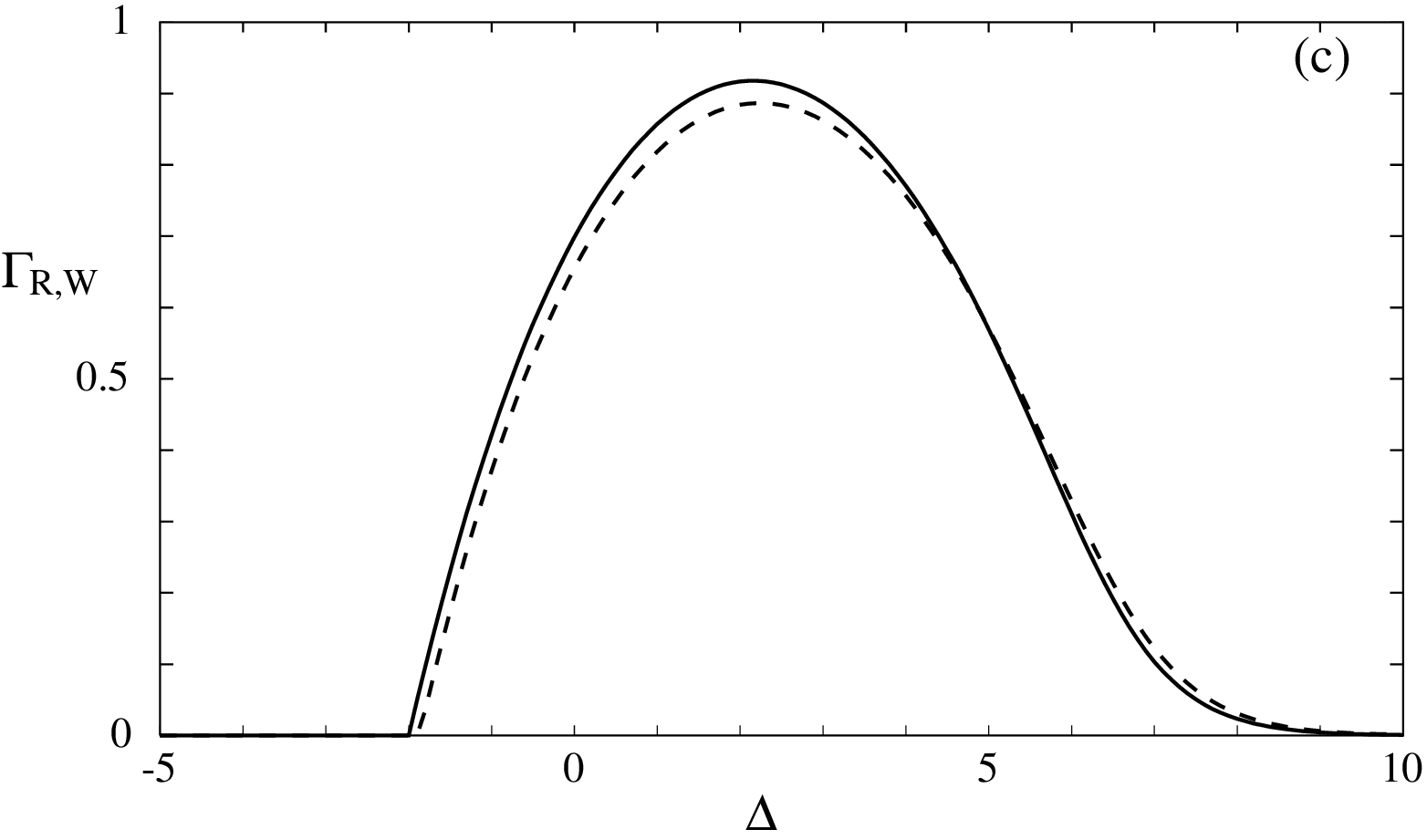,width=8.6cm,clip=}}
\centerline{\psfig{figure=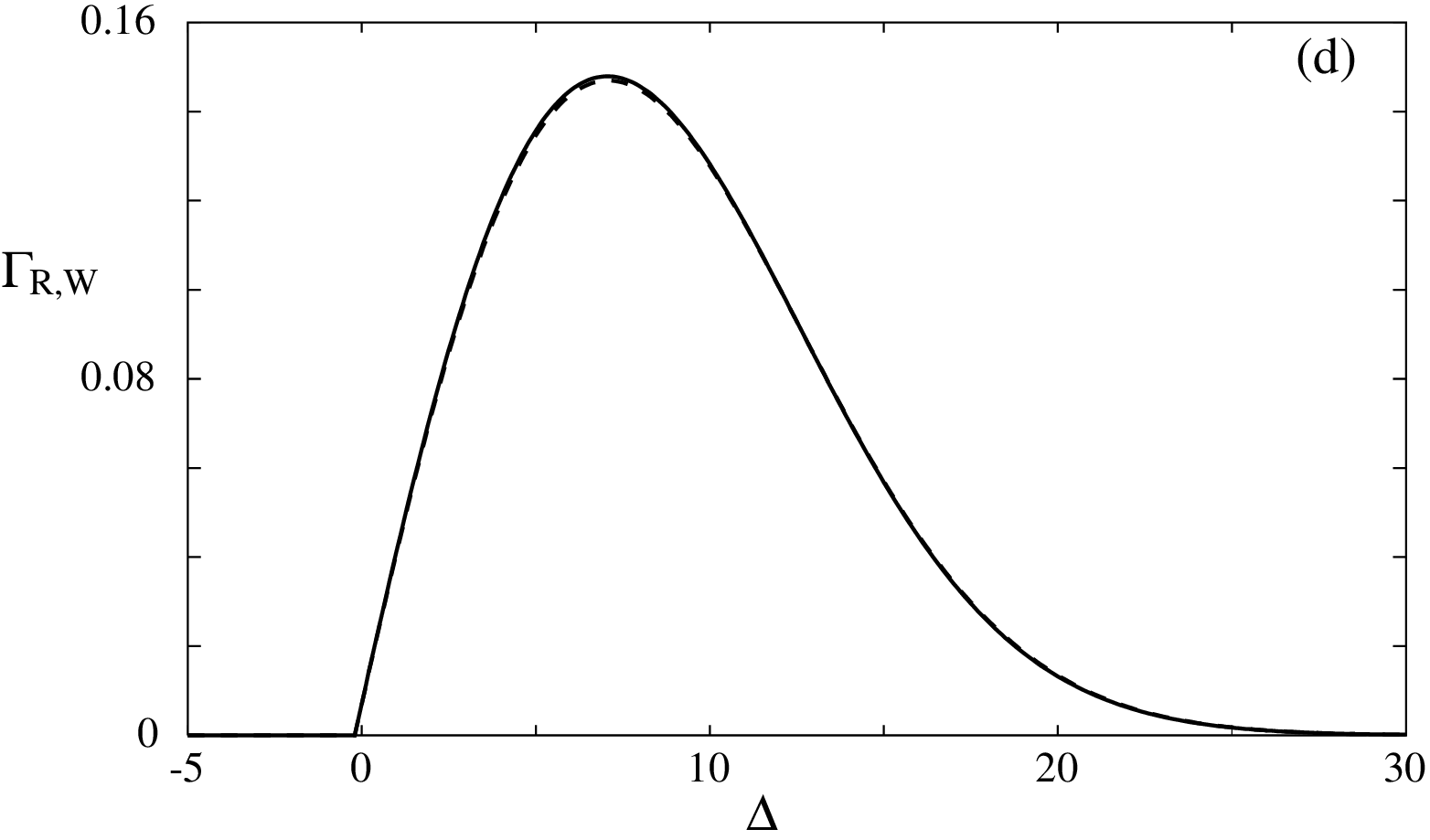,width=8.6cm,clip=}}
\begin{figure}
\caption{Comparison of the exponential growth rate as a function of
pump-probe detuning $\Delta$ between the RAO (solid line) and the
WAO (dashed line) models, for the case $\alpha=10$. (a) shows
the results for $T=0$ (see Sec. III.b),
(b) shows  the case $T=T_R$, (c) shows $T=10T_R$,
and (d) shows $T=100T_R$. We see that the ray atom optics model
gives the correct result only in the limit $T\gg T_R$.}
\end{figure}
\centerline{\psfig{figure=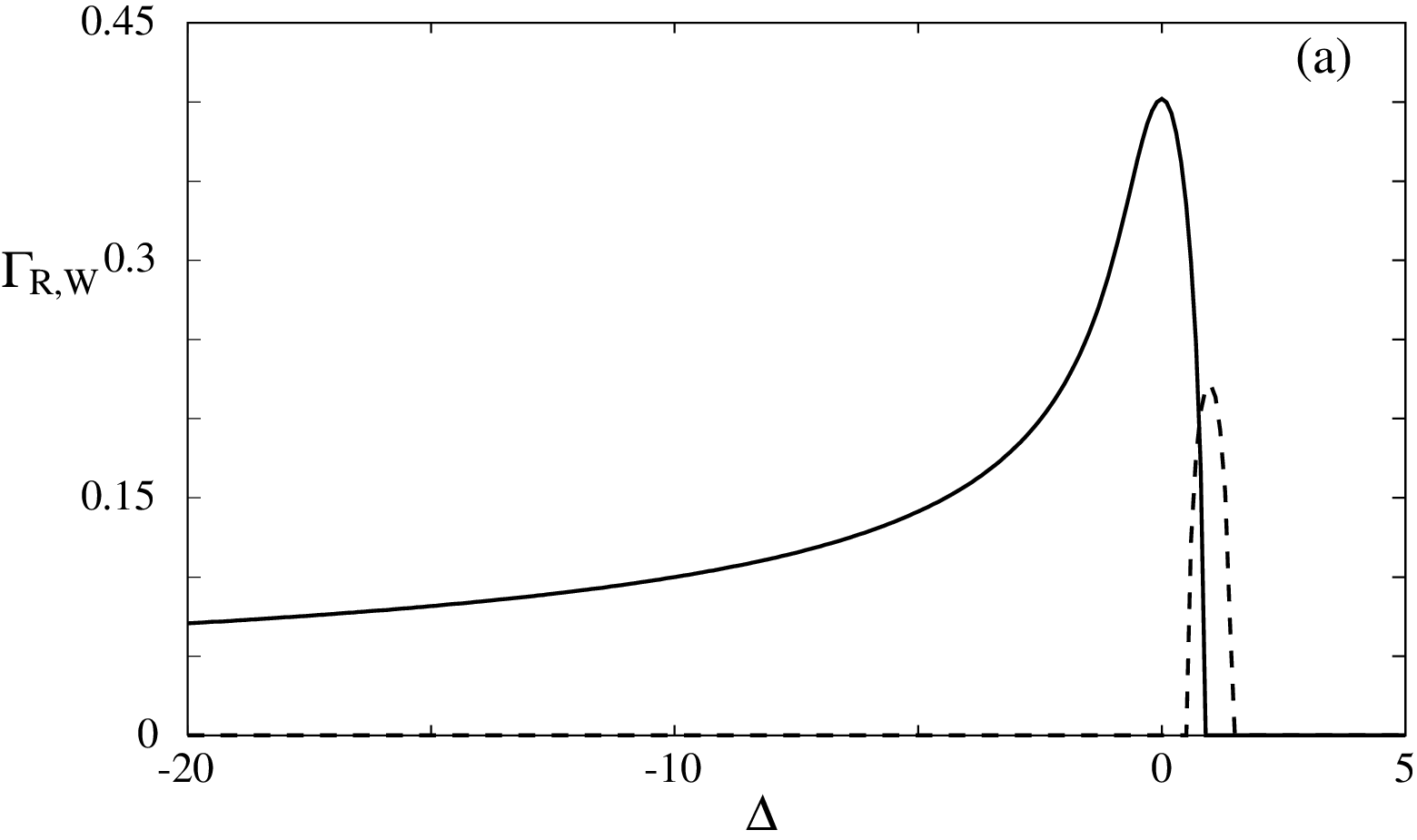,width=8.6cm,clip=}}
\centerline{\psfig{figure=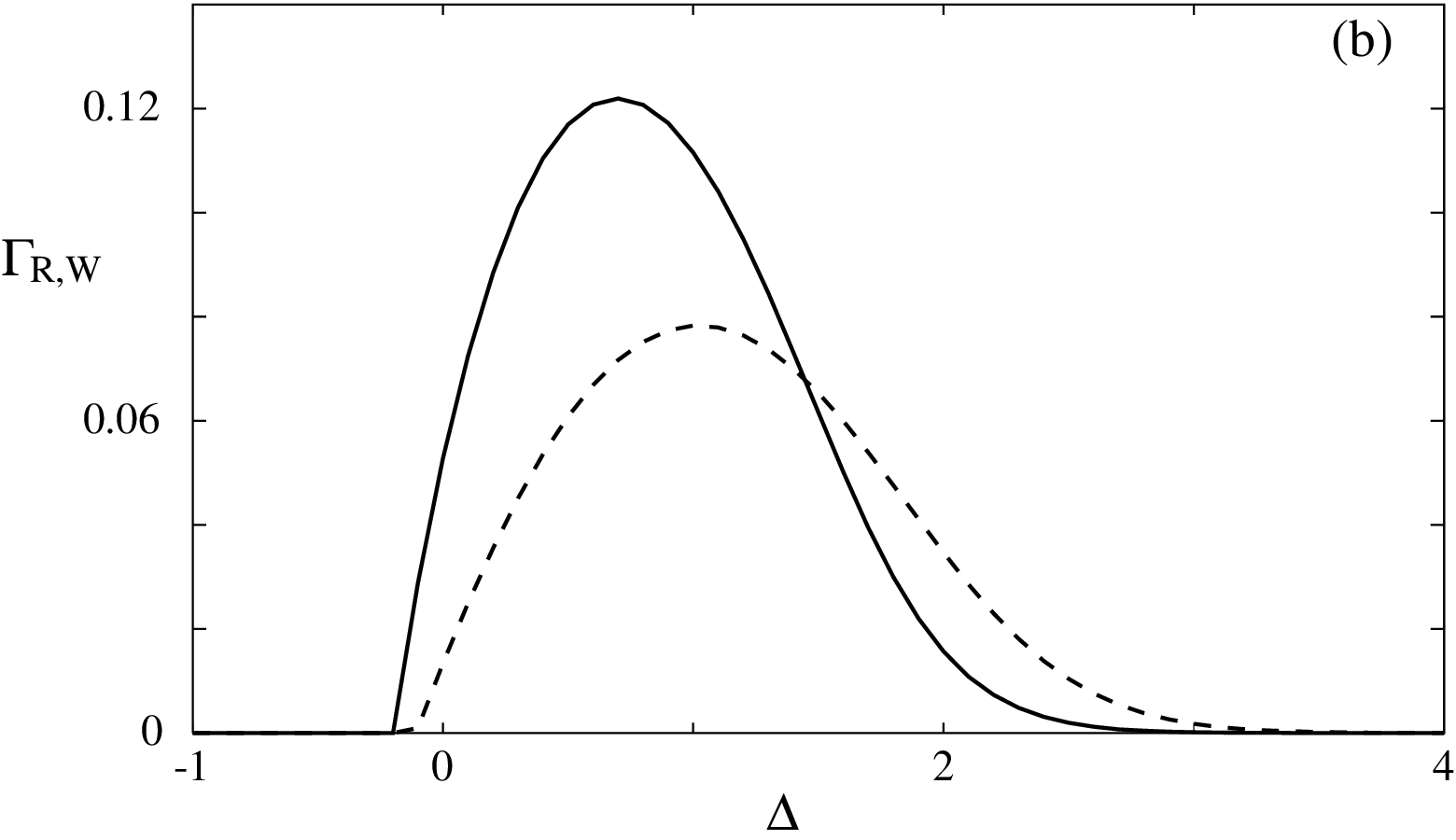,width=8.6cm,clip=}}
\centerline{\psfig{figure=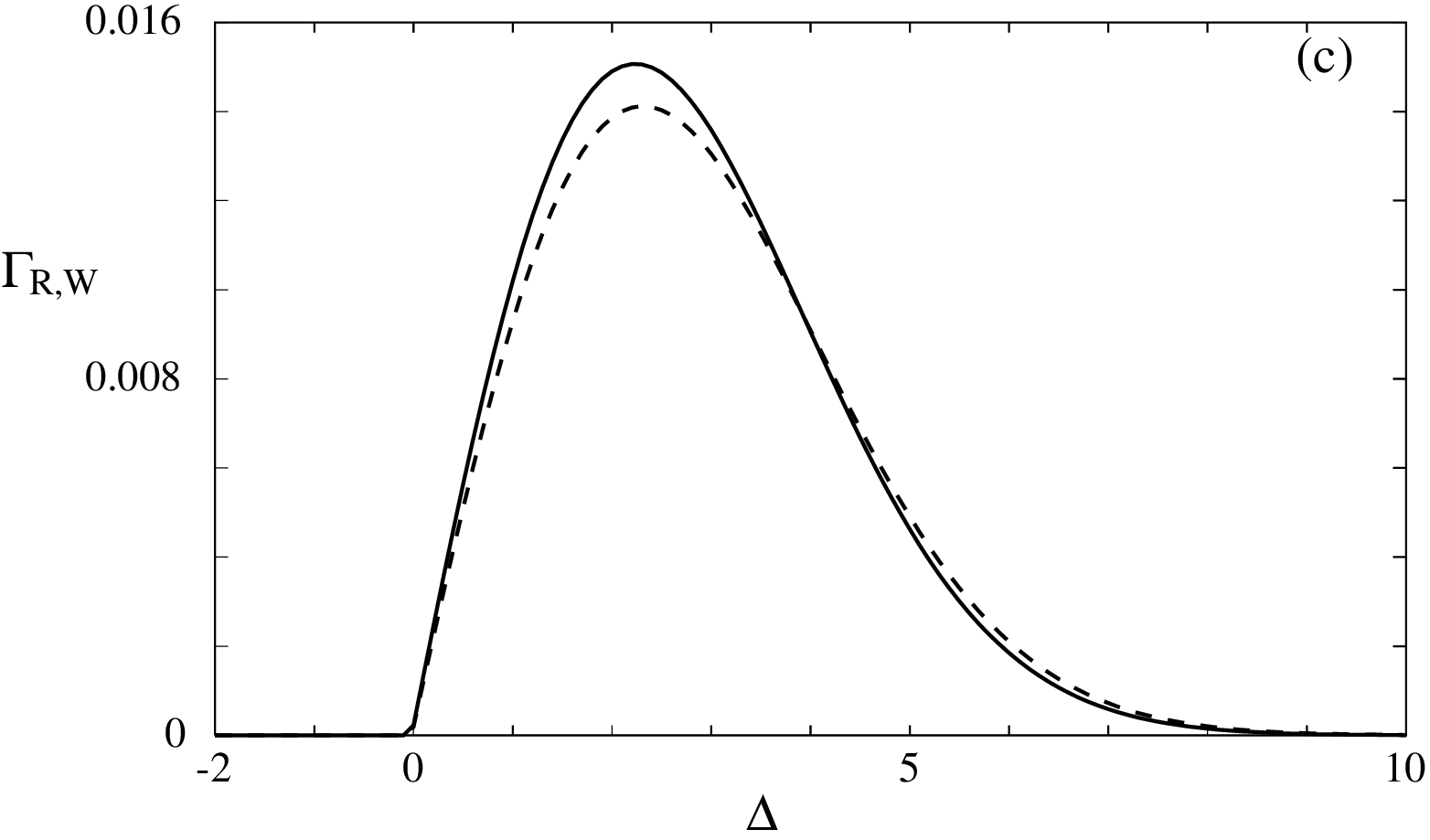,width=8.6cm,clip=}}
\centerline{\psfig{figure=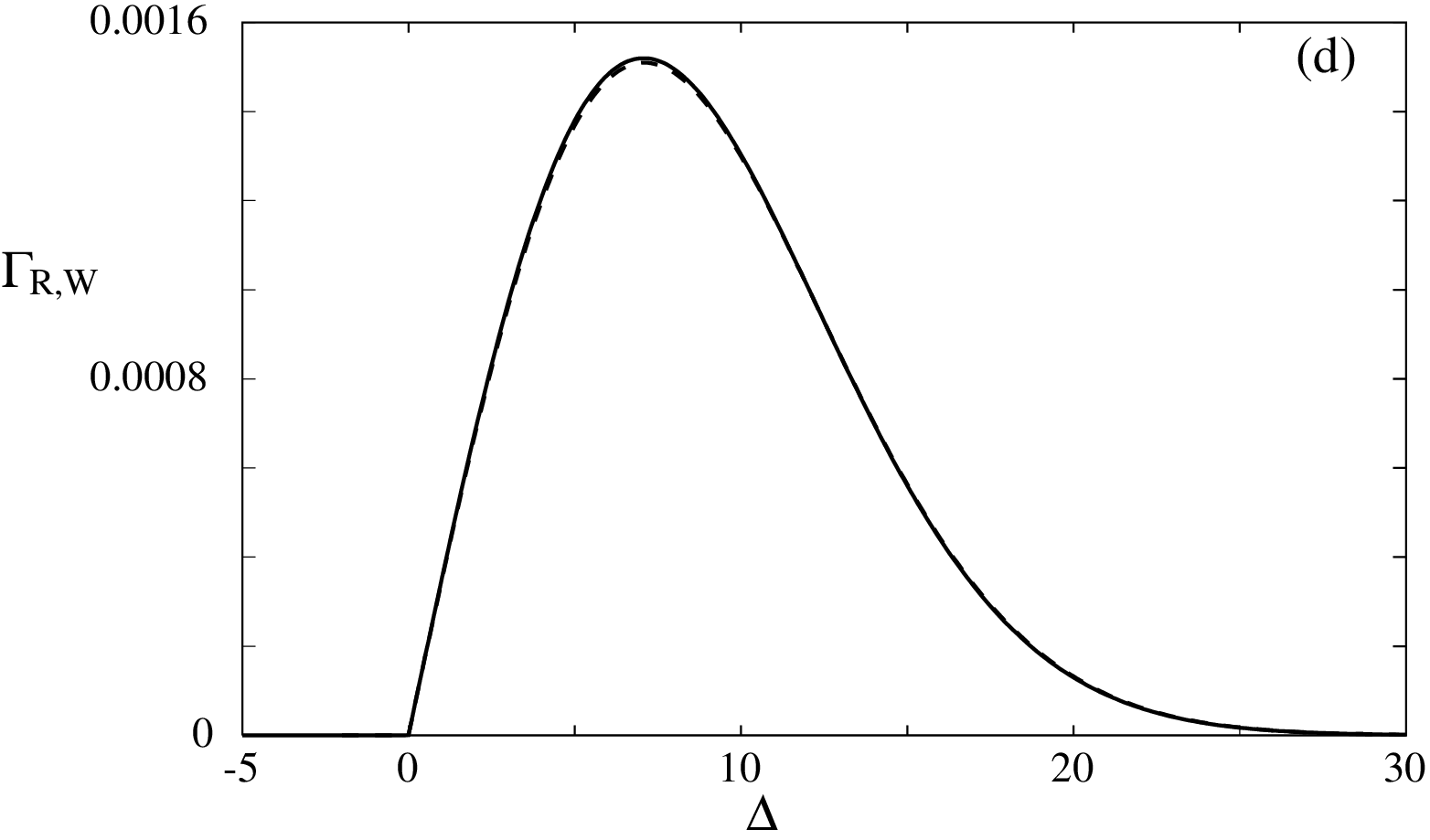,width=8.6cm,clip=}}
\begin{figure}
\caption{Figure 2 is identical to Figure 1, except we have
now taken $\alpha=10^{-1}$.
Since $\alpha$ gives the stength of the bunching
process, when it is small the effects of atomic
diffracton play a larger role, leading to stronger discrepancies
between the predictions of wave atom optics and ray atom optics.
However, we see that the RAO limit, given by $T\gg T_R$, is independent of
$\alpha$.}
\end{figure}
In addition to opposing any increase in the magnitude of $B$, the diffraction
term also modifies its phase, which may upset any phase relation
between $A$ and $B$ which might be required for the collective instability
to occur.

The RAO model only makes accurate predictions at $T=0$ in the limit
$\omega_r\to 0$. Therefore, if we were to increase the mass of the atoms, thus
decreasing $\omega_r$, the behavior at $T=0$ would become more and more
classical. This is because heavier atoms suffer less diffraction
than lighter atoms under the influence of the light fields.
We also note that the correspondence principle states that
quantum mechanics should agree with classical mechanics in the limit
$\hbar\to 0$, which would also cause $\omega_r$ to tend to zero.
These considerations can also be derived from the statement that
the RAO model is valid when $T\gg T_R$, if we note that as $\omega_r\to 0$
the recoil temperature also goes to zero.

In both the RAO and WAO models, the probe field $A$ obeys the 
equation
\begin{equation}
\frac{d}{d\tau}A=i(\Delta A-\alpha B).
\label{zeroA}
\end{equation}
From this equation, together with Eq. (\ref{zeroBrao}) 
we find that the solutions are exponentials with exponents given by
the roots of the cubic equation
\begin{equation}
s^3-i\Delta s^2-i\alpha=0.
\label{cubicrao}
\end{equation}
This is exactly the
``cold-beam'' cubic equation of Bonifacio et al \cite{BonSal95}.
However with the inclusion of atomic diffraction effects, we now see that
the correct ``cold-beam'' cubic equation, derived from Eqs. (\ref{zeroA})
and (\ref{zeroBwao}), is
\begin{equation}
s^3-i\Delta s^2 +s -i(\alpha+\Delta)=0.
\label{cubicwao}
\end{equation}
These equations can also be derived from the Laplace transform 
method of Sec. III, with the substitution $f(k)=\delta(k)$, indicating
a zero temperature momentum distribution.     

From these cubic equations it is possible to determine
the point of transition between the stable and the unstable regimes of the
CARL.
For the RAO model the collective instability occurs provided 
that the threshold condition
\begin{equation}
\alpha>\frac{4\Delta^2}{27}
\label{raothresh}
\end{equation}
is satisfied, and
above threshold the exponential growth rate is given by
\begin{equation}
\Gamma_R=\frac{\sqrt{3}}{2}\left(\frac{\alpha}{4}\right)^{1/3}\left|
(1+\sqrt{C})^{2/3}-(1-\sqrt{C})^{2/3}\right|,
\label{GammaR}
\end{equation}
where $C=1-4\Delta^3/27\alpha$.
For the WAO theory the threshold condition is
\begin{equation}
\alpha>\frac{2}{27}\left[(3+\Delta^2)^{3/2}-9\Delta+\Delta^3\right],
\label{waothresh}
\end{equation}
and above threshold the exponential growth rate is given by
\begin{eqnarray}
\Gamma_W&=&\frac{\sqrt{3}}{2}\left(\frac{\alpha}{4}\right)^{1/3}
\left|\left[(1+\sqrt{D})^2+\frac{4}{27\alpha^2}(1-\Delta^2)^2\right]^{1/3}
\right.\nonumber\\
&-&\left.\left[(1-\sqrt{D})^2+\frac{4}{27\alpha^2}(1-\Delta^2)^2\right]^{1/3}\right|,
\label{GammaW}
\end{eqnarray}
where
\begin{equation}
D=1-\frac{4\Delta}{3\alpha}\left(1-\frac{\Delta^2}{9}\right)
-\frac{4}{27\alpha^2}\left(1-\Delta^2\right)^2.
\label{defD}
\end{equation}

In Figure 3(a) we examine the CARL operating regime,
defined as the region in parameter space where the exponential instability
occurs, at $T=0$ as it would be if Ray Atom Optics were valid.
We contrast this with Figure 3(b) which shows the actual CARL operating
regime at $T=0$, as calculated using Wave Atom Optics. From this figure 
we see that the operating regime of the CARL is drastically reduced at low 
pump intensities and/or atomic densities when the effects of atomic 
diffraction are included.
\centerline{\psfig{figure=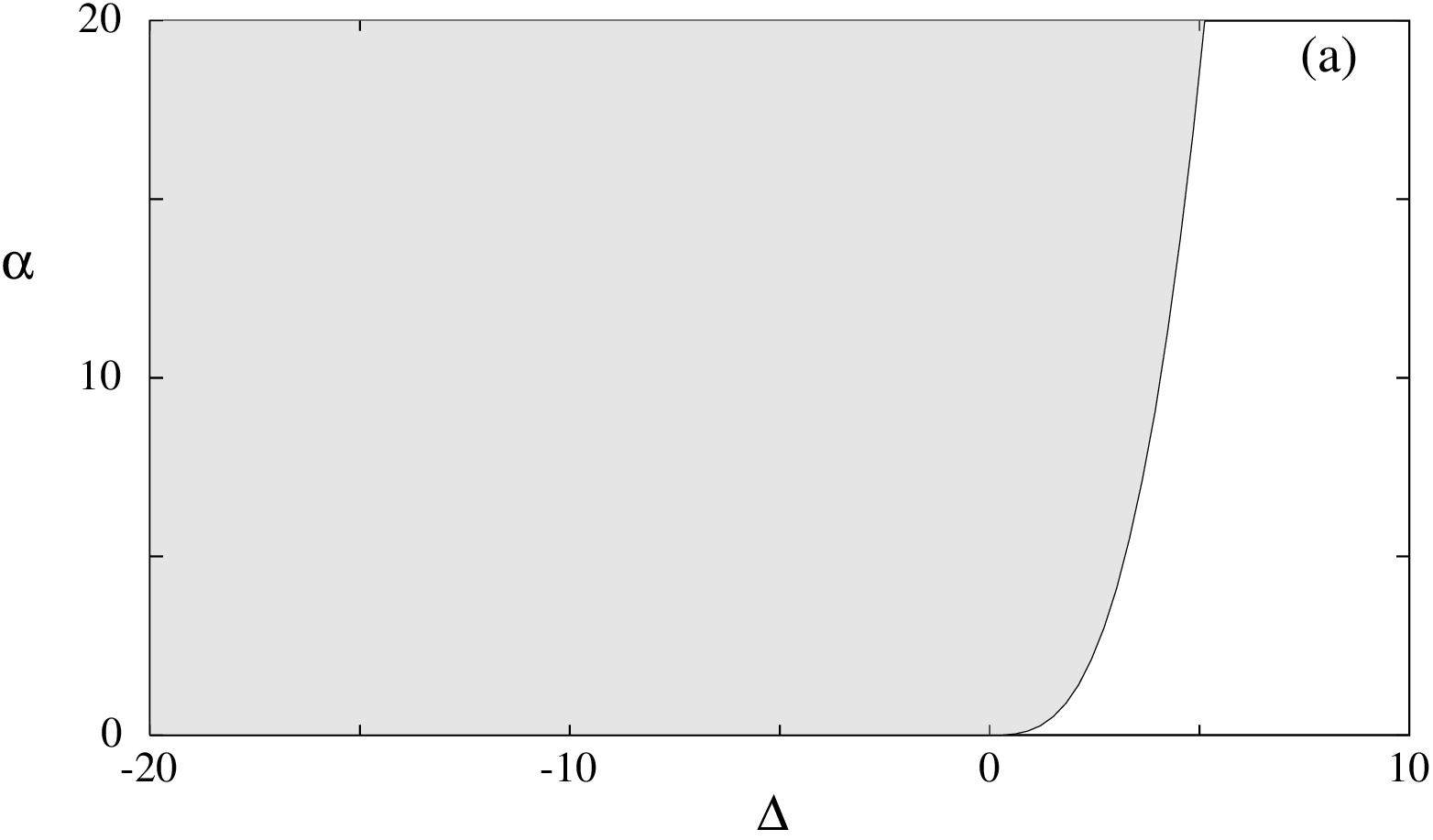,width=8.6cm,clip=}}
\centerline{\psfig{figure=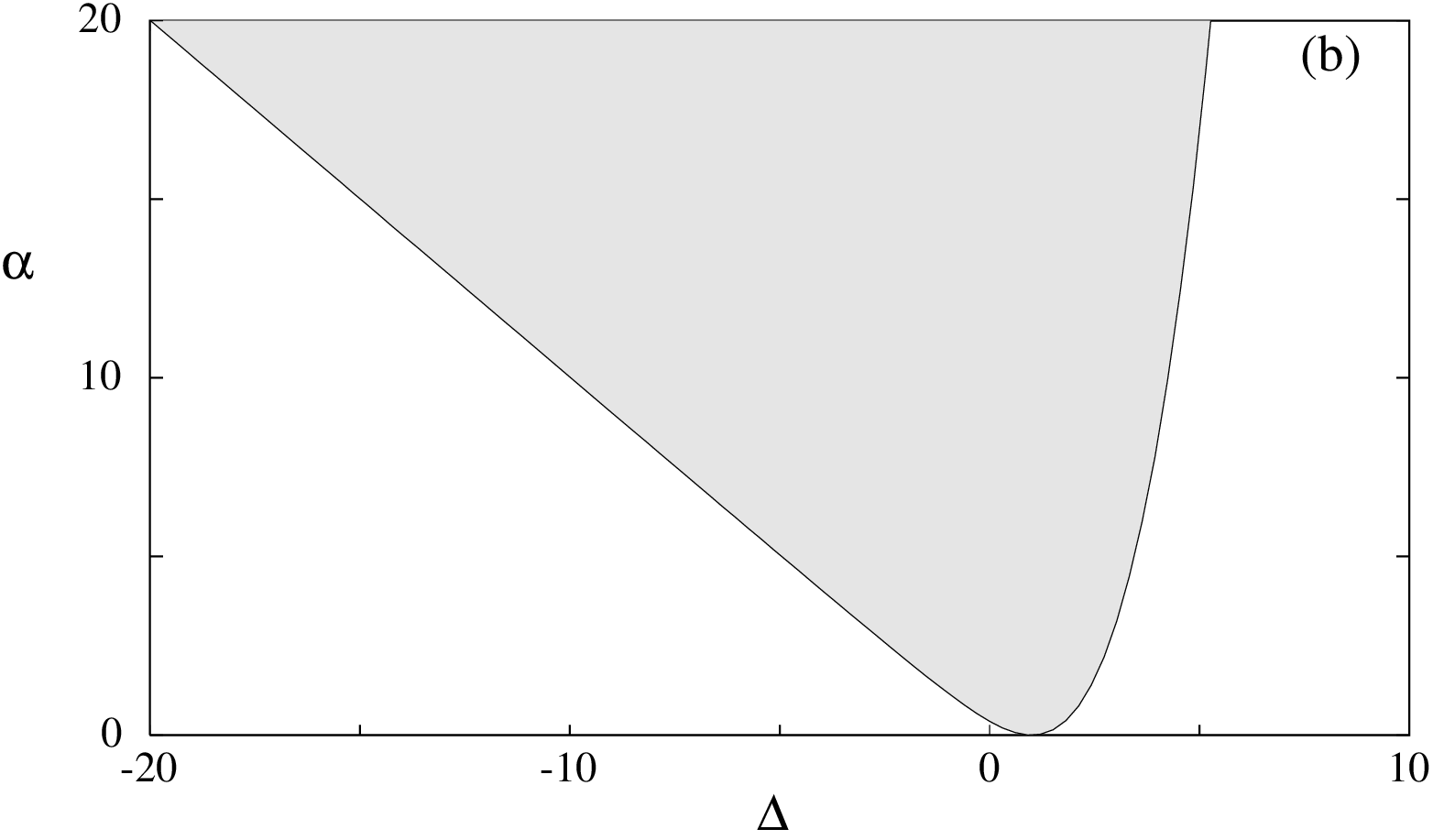,width=8.6cm,clip=}}
\begin{figure}
\caption{The CARL operating regime (shaded region)
as predicted by the RAO model (a), and the actual operating regime (b), as
given by the WAO model.}
\end{figure}

Figure 1(a) compares $\Gamma_R$ with $\Gamma_W$
for the case $\alpha=10$
at $T=0$, and Fig. 2(a) shows the same comparison for $\alpha=10^{-1}$.
We see that atomic diffraction leads to the appearance of a second threshold
below which the collective instability does not occur. From Fig. 2(a) we
see that this second threshold may even be above $\Delta=0$ for low intensities
and/or densities. In fact, the threshold crosses $\Delta=0$ at precisely
$\alpha=2/3\sqrt{3}$.

Figure 2(a) shows that in the limit of weak pump intensities and/or
atomic densities the peak gain for the WAO model tends 
to $\Delta=1$, while that of the RAO model is at $\Delta=0$, 
This result can actually be understood quite simply:
The atomic center-of-mass dispersion curve tells us that the absorption of a
pump photon and the emission of a probe photon by an atom initially at rest
creates an energy defect of $4\omega_r$ due to atomic recoil. This defect can
be compensated by a detuning between the pump and probe, which in dimensionless
units occurs at $\Delta=1$. Therefore, the fact that $\Gamma_W$ is
a sharply peaked function around $\Delta=1$ is simply an expression of
energy-momentum conservation. If we are to take the Ray Atom Optics model 
seriously at $T=0$, then we must concede that we are in the limit where 
$\omega_r\to 0$, therefore, energy-momentum conservation would predict the
maximum of $\Gamma_R$ to occur at $\Delta=0$. In other words, in that limit 
the center-of-mass dispersion curve is flat over the range of a few photon
momenta.

\section{Discussion and outlook}

The main result of this paper is that at low temperatures the behavior of
the CARL is strongly influenced by matter-wave diffraction, which tends to
counteract the atomic bunching and reduces the instability range of the
system. In this temperature range, the CARL presents an experimentally
realizable example of dynamically coupled Schr\"odinger and Maxwell fields.
The present theory quantizes the matter wave, but not the electromagnetic
field. It will be of considerable interest to extend it to regimes
where both fields need to be quantized. An analysis of the density regime
where quantum degeneracy becomes important will also be a fascinating
extension, in particular when two-body collisions are included. This study 
will allow one to investigate to which extent a Bose-Einstein condensate can 
be manipulated and modified in a far off-resonant CARL configuration. An 
intriguing possibility would be to generate in this fashion a coupled 
laser-``atom laser'' system. The study of the coherence properties of this
system will be the object of future investigations. Finally, a comparison 
between bosonic and fermionic CARL systems in the quantum degenerate regime
should also be considered.

\acknowledgements
We have benefited from discussions with R. Bonifacio and L. De Salvo, who
brought the CARL system to our attention. This work is supported in part
by the U.S. Office of Naval Research Contract No. 14-91-J1205, by the 
National Science Foundation Grant PHY95-07639, by the U.S. Army Research 
Office and by the Joint Services Optics Program.

\appendix
\section*{Heisenberg equations of motion for density operators}
The full equations of motion for the expectation values of the
density operators are
\begin{eqnarray}
&&\frac{d}{dt}\rho_{gg}(k,k^\prime)=-\frac{i\hbar}{2m}(k^2-{k^\prime}^2)
\rho_{gg}(k,k^\prime)\nonumber\\
&&+g_1a^\ast_1\rho_{eg}(k-k_1,k^\prime)
+g_2a^\ast_2\rho_{eg}(k-k_2,k^\prime)\nonumber\\
&&+g^\ast_1a_1\rho_{ge}(k,k^\prime-k_1)
+g^\ast_2a_2\rho_{eg}(k,k^\prime-k_2),
\label{drhoggdt}
\end{eqnarray}
\begin{eqnarray}
&&\frac{d}{dt}\rho_{eg}(k,k^\prime)=-i\left[\frac{\hbar}{2m}(k^2-{k^\prime}^2)
+\omega_0\right]\rho_{eg}(k,k^\prime)\nonumber\\
&&+g^\ast_1a_1[\rho_{ee}(k,k^\prime-k_1)
-\rho_{gg}(k+k_1,k^\prime)]\nonumber\\
&&+g^\ast_2a_2[\rho_{ee}(k,k^\prime-k_2)
-\rho_{gg}(k+k_2,k^\prime)],
\label{drhoegdt}
\end{eqnarray}
and
\begin{eqnarray}
&&\frac{d}{dt}\rho_{ee}(k,k^\prime)=-\frac{i\hbar}{2m}(k^2-{k^\prime}^2)
\rho_{ee}(k,k^\prime)\nonumber\\
&&-g_1a^\ast_1\rho_{eg}(k,k^\prime+k_1)
-g_2a^\ast_2\rho_{eg}(k,k^\prime+k_2)\nonumber\\
&&-g^\ast_1a_1\rho_{ge}(k+k_1,k^\prime)
-g^\ast_2a_2\rho_{ge}(k+k_2,k^\prime).
\label{drhoeedt}
\end{eqnarray}

%\centerline{\psfig{figure=fig1a.ps,width=8.6cm,clip=}}
%\centerline{\psfig{figure=fig1b.ps,width=8.6cm,clip=}}
%\begin{figure}
%\caption{Comparison of the linear growth rate versus pump-probe 
%detuning between the RAO and WAO regimes.  For each curve, 
%the value of the product $\alpha\beta$ is given. Fig. 1a shows $\Gamma_R$, 
%while Fig. 1b shows $\Gamma_W$.}
%\end{figure}

\end{document}